\documentclass[12pt]{article}
\usepackage{epsf}
\def\beq{\begin{eqnarray}}   \def\eeq{\end{eqnarray}}

\begin{document}
\begin{titlepage}
\begin{flushright}
ITEP-TH-45-01\\
TPI-MINN-39/01  \\
UMN-TH-2021\\




\end{flushright}

\vspace{0.6cm}

\begin{center}
\Large{{\bf Reduced ${\cal N}=2$
Quantum Mechanics: Descendants of the  K\"ahler
Geometries}}

\vspace{1cm}

 A. Losev$^{1,2)}$ and M. Shifman$^{1)}$
 \end{center}
\vspace{0.3cm}

\begin{center}
$^1$  {\em Theoretical Physics Institute, Univ. of Minnesota,
Minneapolis,
       MN 55455}

$^2$ {\em Institute of Theoretical and Experimental Physics, Moscow
117259,
Russia}$^\dagger$
 \end{center}

\vspace{1.5cm}

\begin{abstract}
We discuss an ${\cal N}=2$ quantum mechanics with or without a 
central charge.
A representation is constructed with the number of
bosonic degrees of freedom less that one half of the
fermionic degrees of freedom. We suggest a systematic method of reducing
the bosonic degrees of freedom called ``dynamical reduction."
Our consideration opens a problem of a  
general classification of nonstandard representations of
${\cal N}=2$ superalgebra.

\end{abstract}

\vspace{5cm}

\begin{flushleft}
--------------------------------\\
$^\dagger$ Permanent address
\end{flushleft}
\end{titlepage}

\section{Introduction}
\label{sec1}

Solution of many physical problems lead to supersymmetric quantum
mechanics (SQM) \cite{EW}. Suffice it to mention quantization of moduli 
of
supersymmetric solitons. On the other hand many geometrical concepts
(such as K\"ahler and hyper-K\"ahler structures) are in one-to-one
correspondence with extended-SUSY quantum mechanics.
Therefore, it is natural to expect that,
studying novel examples of extended-SUSY quantum mechanics,
one can reveal new geometries of interest.

Quantum mechanics with extended supersymmetry was
considered previously both  in the physical and mathematical
literature \cite{qmes,A,B,exexmb,C,D,E}. The most common strategy for obtaining
such systems is dimensional reduction from $D=2$ or
$D=4$. In four dimensions minimal supersymmetry
has four supercharges. Upon reduction to $D=1$
one obtains a quantal system with four supercharges,
i.e. ${\cal N}=2$. The situation
most extensively studied in the literature corresponds to
\beq
\nu_F = 2\nu_B\, ,
\label{msone}
\eeq
where $\nu_{B,F}$ stands for the number of bosonic (fermionic)
degrees of freedom, respectively.
We will refer to this pattern as standard.  A few exotic examples with
$\nu_B > \nu_F/2$ were analyzed too.
In this work we address the problem of constructing and analyzing
systems with $$\nu_B < \nu_F/2\,.$$ We will suggest a regular method
which may be called a {\em dynamical reduction} of  bosonic variables.

The general representation of the superalgebra  with four supercharges
(SQM$_2$)
\beq
\{Q_i\,, Q_j\}
=2\delta_{ij}H\,,  \quad i,j=1,2,3,4\,,
\label{stalg}
\eeq to be
investigated below, is
\beq
Q_i = \psi^a\,  e^\mu_{i;a}\, \frac{\partial}{\partial X^\mu} +
\eta_{i;abc}
\,
\psi^a \psi^b \psi^c
\label{repre}
\eeq
where
\beq
\{\psi^a \,, \psi^b \} = 2\delta^{ab}\, ,
\eeq
$X^\mu$ and $\psi^a$ are bosonic and fermionic
coordinates, respectively.
As well-known, extended supersymmetry implies $R$ symmetry. Let
$G_A$ be a generator of $R$ symmetry, i.e.
\beq
[G_A, Q_i]= U_{A;i}^j Q_j\,,
\eeq
where $U_{A;i}^j $ are constant matrices. The generators $G_A$
form the $R$ algebra. In the standard situation it is SU(2).

Our main results are:

\vspace{0.2cm}

(I) Obtaining representation (\ref{repre}) of SQM$_2$ with

\hspace{1cm} (i) $\mu=1,\,\,\, a=1,2,3,4$, i.e. $\nu_F = 4 \nu_B$;

\hspace{1cm} (ii) overextended $R$ symmetry so(4) =
su(2)$+$su(2);

\hspace{1cm} (iii) a non-trivial Hamiltonian;

(II) Modifying representations obtained in (I)  to incorporate
central charges (see Eq. (\ref{L8})):

\hspace{1cm} (i) the central charge has a geometrical meaning;

\hspace{1cm} (ii) we found a  mirror-like symmetry;

(III) Our procedure of dynamical reduction explains, in part,
points (I),  (II).

\vspace{0.2cm}

There is a certain overlap between the results presented here and
those obtained previously within different approaches. In particular,
the issue of constructing superalgebras of a more general form
was addressed in Ref. \cite{B}. Moreover, the overextended
$R$ symmetry naturally appears in a superfield approach of 
Ref. \cite{D} which explains also why it disappears
upon inclusion of the central charge \footnote{We have learned of the existence
of this illuminating work only after the submission of our paper to hep-th.
We are grateful to E. Ivanov,  S. Krivonos and A. Pashnev for drawing our attention
to
Ref. \cite{D}.}. The inclusion of the central charge was studied in Ref. \cite{E}. Our
approach of  {\em dynamical reduction} of bosonic coordinates
has its own merits; it is transparent, has  a clear-cut  geometrical interpretation
and reveals the geometric connection 
 of the central charges.

The organization of the paper is as follows.
In Sec. 2 we present a new  solution of the SQM$_2$ equations,
and explain that its $R$ symmetry is larger than for the standard
solutions.
In Sec. 3 
we review some aspects of 
centrally extended superalgebra with four supercharges, SQM$_{2,Z}$.
The solution presented in Sec. 2 is generalized to include the central
 charge. It is noted that the quantal system thus obtained,
SQM$_{2,Z}$ possesses a surprising mirror-like symmetry.
We investigate the nonrelativistic limit of this system.
In Sec. 4 we explain how such systems can be obtained through a
dynamical reduction of the standard ${\cal N}=2$
quantum mechanics on the K\"ahler manifolds. We present a
 clear-cut geometrical formulation.
Finally, in Sec. 5 we outline some
issues to be studied in the future.

\section{A Nonstandard Example with Overextended
$R$ Symmetry}

\subsection{The example}

The realization of the algebra (\ref{stalg}) to be considered here,
 is built
on one bosonic variable
$X$, and 4 fermionic,
\beq
\{ \psi_i , \psi_j \} =2 \delta_{ij},\quad  i,j=1,\ldots ,4\,.
\label{Lthree}
\eeq
For what follows we will introduce also $\psi_5$,
\beq
 \psi_{5}=\psi_{1} \psi_{2} \psi_{3} \psi_{4}\,,\quad \{ \psi_i ,\psi_5\}
=0\,,\,\,\,  (i=1,\ldots ,4),\quad \quad ( \psi_{5})^2 =1\, .
\label{Lfour}
\eeq

The representation  we want to construct
depends on one function of one real
variable $f(X)$,
and is given by
 \beq
  Q_{j}&= & i \psi_{j} \frac{\partial}{ \partial X}+  i f(X)  \psi_j
\psi_{5} \,,
\quad j=1,\ldots, 4\,,
 \nonumber\\[0.2cm]
 H&= & - \frac{d^2}{dX^2} + (f(X))^2 - \psi_5  \frac{df}{dX}\,.
\label{Lfive}
 \eeq

Since the algebra (\ref{Lthree}) is nothing but the
Clifford algebra, it is realized by four-by-four (Euclidean)
$\gamma$ matrices.
 Let $S$ be the space of the four-component spinors,
 and $S_{\pm}$ the spaces
of chiral and antichiral spinors,
$$\psi_5 S_{\pm}=\pm S_{\pm}\,.$$

If we look only at the Hamiltonian, and, for a short while,
forget about the supercharges, we will immediately  see that this
 system
is a combination of
 two {\em identical} decoupled   Witten's Hamiltonians, each of them
presenting ${\cal N}=1$
supersymmetric quantum mechanics. Therefore, the system
(\ref{Lfive}) has  the following obvious properties.

(i) Any excited (nonvacuum) state has degeneracy  equal to 4,
i.e. the dimension of the supermultiplet is four.

As for the ground state, there are
several possibilities.

(ii)
The bosonic target space  is compact (i.e. the coordinate  $X$
lives on a circle $S_1$); and the period
\beq
\Pi = \int_{S_{1}} f(X) dX
\label{Lsix}
\eeq
vanishes. Then SUSY is unbroken, there are four zero-energy states.
Two are  in the space of the chiral spinor fields, and two in the space
of the antichiral spinor  fields,
\beq
\Psi_{a,\pm} (X) = s_{a,\pm}\exp\left( \pm  \int^{X} f(T)dT\right)\, , 
\quad
a=1,2\,,
\label{Lseven}
\eeq
where $s_{+,a} \,\, (s_{-,a})$ denote the basis in the space of the
(anti)chiral spinors.

(iii) The bosonic target space  is compact and the period
(\ref{Lsix}) does not vanish.
 Then SUSY is broken, there are no zero-energy states.
The degeneracy of the (nonsupersymmetric) ground state is four.

If the bosonic target space is noncompact, then
there are two possibilities:

(iv)  neither $\Psi_{+}$, nor $\Psi_{-}$ (see Eq. (\ref{Lseven}))
are  normalizable; SUSY is broken, there are no zero-energy states.

(v) either $\Psi_{+}$, or $\Psi_{-}$ is normalizable;
SUSY is unbroken, the degeneracy of
the  zero-energy state is two.

\subsection{The overextended $R$ symmetry}

As well-known, in the standard (K\"ahler)
case the $R$ symmetry of ${\cal N}=2$
quantum mechanics is SU(2). In the K\"ahler  sigma models
(without superpotential) this symmetry is known as the Lefshetz
SL(2) symmetry.

In the example constructed above the $R$ symmetry is larger, it is
SO(4). The supercharges $Q_i\quad (i=1,2,3,4)$ form a representation
$\underline{4}$ of this SO(4). Indeed, the generators of SO(4) are
\beq
G_{ab} =\frac{1}{2}[\psi^a\,,\psi^b]\, .
\eeq

Note that in the standard realization, even on the flat metric complex
plane, the
four supercharges form two doublets of SU(2), which cannot be 
transformed
one
into another by $R$ symmetry, since one pair of the supercharges 
contain
$\partial /\partial z$ and another pair $\partial /\partial \bar z$. (Here
$z$ is the
complex coordinate on the plane).

Our SO(4) $R$ symmetry
should not be confused with SU(4) that acts on
{\em nonvacuum} states in the standard realization of ${\cal N}=2$
SUSY, with the generators
\beq
[Q_i\,,Q_j]\, \frac{1}{2H}\,.
\eeq

\section{Centrally Extended ${\cal N}=2$ Superalgebra}

The general ${\cal N}=2$ algebra with the central charges
 can be
written as follows (in the real notations)
\beq
(Q_1)^2 &=& (Q_2)^2 = H-Z \,,\qquad (Q_3)^2 = (Q_4)^2 = H+Z \,,
\nonumber \\[0.2cm]
\{Q_1 , Q_3\}&=&
\{Q_2,  Q_4\}= -2 P\,,
\label{one}
\eeq
with all other anticommutators vanishing. We assumed that the central
charge $Z$ in Eq. (\ref{one}) is real (and positive).
This can be always achieved by an appropriate phase
rotation in the definition of the  supercharges. Here $H$ is the
Hamiltonian,
$P$ is the momentum operator.
Restricting ourselves to the sector with the vanishing spatial
momentum ${P}=0$ we obtain the following algebra
\beq
\{ Q_i, Q_j\} =2 \left(\begin{array}{cccc}
H-Z &0&0&0\\[0.1cm]
0&H-Z&0&0\\[0.1cm]
0&0&H+Z&0\\[0.1cm]
0&0&0&H+Z
\end{array}
\right)_{ij}\,,\qquad i,j= 1,...,4\,.
\label{two}
\eeq
We will refer to it as (SQM$_{2,Z}$ for short).

The simplest realization of the algebra (\ref{two}),
which presents a straightforward generalization of Witten's quantum
mechanics with two supercharges
\cite{EW}, was suggested in \cite{3},
\begin{eqnarray}
Q_1 &=& \psi_1\, p +W'(x)\psi_2\,,\quad Q_2 = \psi_2\, p -W'(x)\psi_1\,
\nonumber\\[0.2cm]
Q_3 &=& \psi_4\left(2Z +p^2 +(W')^2-i\psi_1\psi_2W''
\right)^{1/2}\, ,\nonumber\\[0.2cm]
Q_4 &=&-\psi_3 \left(2Z +p^2 +(W')^2-i\psi_1\psi_2W''
\right)^{1/2}\,,
\label{rsvv}
\end{eqnarray}
where $W$ is a superpotential depending on one bosonic variable
$x$, and $p = -i d/dx$.
Moreover,
\beq
\{\psi_i , \psi_j\} = 2\delta_{ij}\,.
\eeq

This realization is natural from the standpoint of the
nonrelativistic expansion.
In the  nonrelativistic limit the first
two supercharges are small ($p\sim W' \sim \beta$), and so is $H-Z$
(which is proportional to $\beta^2$).
The last two supercharges are large, to the leading order in $\beta$
\beq
Q_3 = \psi_4 \sqrt{2Z}\,,\quad Q_4 = -\psi_3 \sqrt{2Z}\,.
\label{dopseven}
\eeq
The Hamiltonian can be obtained by squaring $Q_1$ or $Q_2$,
\beq
H - Z =  p^2 +(W')^2-i\psi_1\psi_2W''\, .
\label{dopone}
\eeq

\vspace{0.2cm}

The realization (\ref{rsvv}) is
nonlinear in
$p$, however (see \cite{3} for further details).
It is evident that the fermionic variables $\psi_{3,4}$ are factored out;
$H-Z$ does not depend on them;
 the supercharges $Q_{3,4}$ play no dynamical role apart from ensuring
the proper dimension of the supermultiplets.
In essence, the system described by (\ref{rsvv})
reduces to Witten's quantum
mechanics with two supercharges.

Our task is to explore other realizations of ${\cal N}=2$ centrally 
extended
SUSY algebra -- linear in $p$.

\subsection{The  Centrally Extended  Reduced Quantum Mechanics}

So far we considered a superalgebra with four supercharges and no
central extension. In the presence of central charges
the general representation of the supercharges takes the form
\beq
Q_i = \psi^a\,  e^\mu_{i;a}\, \frac{\partial}{\partial X^\mu} +
\eta_{i;abc}
\,
\psi^a \psi^b \psi^c +\zeta_{i;a}\psi^a \,,
\label{cerepre}
\eeq
i.e. one  adds terms linear in $\psi$
to the supercharges, and obtain  an SQM$_{2,Z}$ algebra. More 
concretely,
\beq
Q_1& = &  i \psi_{1} \frac{\partial}{ \partial X}+
 i f(X)  \psi_1 \psi_{5} + \psi_2 \left( -v g(X) + \frac{\omega}{
g(X)}\right) \,,
  \nonumber \\ [0.2cm]
Q_2& = &  i \psi_{2} \frac{\partial}{ \partial X}+
 i f(X)  \psi_2 \psi_{5} + \psi_1\left( +v g(X) - \frac{\omega}{
g(X)}\right) \,,
   \nonumber \\ [0.2cm]
Q_3& = &  i \psi_{3} \frac{\partial}{ \partial X}+
 i f(X)  \psi_3 \psi_{5} + \psi_4 \left( +v g(X) + \frac{\omega}{
g(X)}\right) \,,
   \nonumber \\ [0.2cm]
Q_4& = &  i \psi_{4} \frac{\partial}{ \partial X}+
 i f(X)  \psi_4 \psi_{5} + \psi_3 \left( - v g(X) -\frac{\omega}{
g(X)}\right) \,,
\label{L8}
\eeq
where $v$ and $\omega$ are constants, while the functions $f(X)$ and 
$g(X)$
are related as follows:
\beq
- \frac{1}{2}\frac{d}{dX} \log g =f(X)\, .
\label{L9}
\eeq
The central charge is  equal to
\beq
Z=2\, v\,  \omega \,.
\label{Lten}
\eeq
Equation (\ref{L9}) implies, in particular, that if the target space
is compact and the period $\Pi\neq 0$ (see Eq. (\ref{Lsix}))
the central extension of the type (\ref{L8}) is impossible,
additional terms linear in $\psi$'s cannot be introduced.

As we will see below, the realization (\ref{L8})
is obtained by a certain reduction from the standard K\"ahlerian
formulation; therefore, we will refer to it as to {\em reduced
K\"ahler-related}.

The Hamiltonian corresponding to the given realization of
SQM$_{2,Z}$ is
 \beq
 H & = & - \frac{d^2}{dX^2} +
\left(-vg(X) +\frac{\omega}{g(X)} +i\psi_3\psi_4\,  f(X)
\right)^2  \nonumber \\[0.3cm]
 & +&i \psi_1\psi_2\frac{d}{dX} \left(-vg(X) +\frac{\omega}{g(X)}
+i\psi_3\psi_4\, f(X)
\right)+2v\omega\, .
\label{Leleven}
 \eeq
Alternatively this Hamiltonian can be rewritten as
\beq
 H & = & - \frac{d^2}{dX^2} + v^2\, g^2+\frac{\omega^2}{g^2} + f^2
 \nonumber \\[0.3cm]
 & +&2i \psi_3\psi_4 \, f\left(-v\, g +\frac{\omega}{g}
\right) +2i \psi_1\psi_2 \, f\left(v\, g +\frac{\omega}{g}
\right) -\psi_5 \, \frac{df}{dX}\,,
\label{Ltwelve}
 \eeq
where the first and the second line represent the
boson and fermion terms, respectively. Putting $v=\omega =0$
returns us to Eq. (\ref{Lfive}).

As was mentioned above, the algebra (\ref{Lthree}) can be realized
by the (Euclidean) $\gamma$ matrices. One can choose them in such a 
way
that $\psi_1\psi_2$, $\psi_3\psi_4$ and $\psi_5$ are all diagonal,
e.g.
\beq
i \psi_1\psi_2&=&{\rm diag} (-1,1,1,-1)\,,\quad
i \psi_3\psi_4={\rm diag} (-1,1,-1,1)\,,\nonumber\\[0.2cm]
\psi_5 &=&{\rm diag} (-1,-1,1,1)\, .
\eeq

It is not difficult to  construct the  wave functions of the ground state,
assuming for definiteness that $Z>0$.
To this
end one  multiplies the supercharge $Q_1$ by $\psi_1$
and requires
\beq
\psi_1 Q_1 |\Psi_0\rangle = 0\,.
\eeq
Consider the basis in $S_{\pm}$ such that $\psi_1 \psi_2$ is diagonal
in this basis. The formal solutions to this equation
have the form
\beq
\Psi=S^{(a)}\left( g(X)\right)^{\varepsilon_a/2} \exp\left\{ \delta_{a}
\int\left(  v g(X)-\frac{\omega}{ g(X)}\right) dX \right\}\, ,
\label{doptwo}
\eeq
here $\varepsilon_a$ and $\delta_a$ are $\pm 1$ . They are
common eigenvalues of  the operators $i\psi_1 \psi_2$
and $i\psi_3 \psi_4$, while  $S^{(a)}$ are the corresponding 
eigenvectors.
For instance, if
\beq
S^{(1)} =\left(\begin{array}{c}
1\\0\\0\\0\end{array}\right)\,,
\eeq
then
\beq
\varepsilon_1 = -1,\quad \delta_1 = 1\,.
\eeq
Whether the formal solution is the actual ground state
with $E_0 = Z$ depends on normalizability. Assume that the target space 
is
noncompact. It is natural to start from the case when $g(X)$ neither
vanishes
nor  becomes infinite at $-\infty < X<\infty $.  Then the answer depends 
on
the number of zeros of
\beq
 F(X) \equiv v g(X)-\frac{\omega}{ g(X)}\, .
\eeq
The point where $F(X)=0$ is the classical ground state.
If this number is odd, two $E_0 = Z$ states of the type
(\ref{doptwo}) are normalizable. They correspond to a specific choice
of $\delta_a$ and $\varepsilon_a = \pm 1$. If the number of zeros is 
even
(or
no zeros at all), supersymmetry is broken, and the ground state
has $E_0>Z$. All multiplets with $E_0>Z$ have dimension four.
This is a quantum-mechanical manifestation of the
multiplet shortening at $E_0 =Z$.

If
$g(X) =0$ or $g(X) =\infty$ at some finite $X=X_*$,
 the problem actually splits in two, one defined on the interval
$(-\infty , X_*)$, another on the interval $(X_*, \infty )$.
They can and must  be considered separately.

Unlike the Hamiltonian (\ref{dopone})
where two fermion variables are factored out,
in the case at hand, see Eq. (\ref{Leleven}),  there is no factorization --
 spin is  entangled with the coordinate motion.

\subsection{``Mirror" symmetry}

Note, that the realization of SQM$_{2,Z}$  we have constructed
has a strange ``mirror" symmetry.
Namely, if one  interchanges
\beq
&& \omega \rightarrow v\, , \quad v \rightarrow \omega\, ;
\nonumber\\[0.2cm]
&& g(X) \rightarrow \frac{1}{g(X)}\,,\qquad(\mbox{then}\,\,\,
f\to -f)\,,
\label{dop3}
\eeq
this leaves
the purely bosonic part of the  Hamiltonian (the first line in Eq.
(\ref{Ltwelve})) invariant. Now one can  promote this symmetry of the
bosonic part of the Hamiltonian to the symmetry of the supercharges,
and the fermionic part of the Hamiltonian. To this end one must
supplement Eq. (\ref{dop3}) by
\beq
\psi_{1} &\rightarrow& \psi_2\,,\quad  \psi_2 \rightarrow \psi_1\,;
\nonumber
\\[0.2cm]
\psi_{3} &\rightarrow &\psi_4\,,\quad \psi_4 \rightarrow -\psi_3\,.
\label{dop4}
\eeq
Under the combined action of (\ref{dop3}) and (\ref{dop4})
\beq
Q_1\leftrightarrow Q_2\,,\quad Q_3\to Q_4\,,\quad Q_4\to - Q_3\,.
\eeq

Summarizing, the ``mirror" symmetry is an intrinsic
 property of the reduced K\"ahler-related SQM. A mirror
symmetry was also observed by Ivanov {\em et al.} \cite{D}
within their superfiled approach.

\subsection{Nonrelativistic limit}

It is instructive to examine the nonrelativistic limit of
the mirror-symmetric
realization (\ref{L8}). In our convention the particle mass is $1/2$;
we will adjust the central charge correspondingly, putting $Z=1/2$.
The simplest way to proceed to the nonrelativistic limit is as
follows
\beq
v\to \frac{1}{2}\,,\quad \omega \to \frac{1}{2}\,;\nonumber\\[0.2cm]
g(X) \to 1+\beta (X)\, ,
\eeq
where it will be assumed that
$$
|\beta (X) |\ll 1\,,\quad \beta (X) \sim \beta \,\,\,(\mbox{velocity})
$$
is the expansion parameter. Then
$$
d/dX \sim \beta , \quad\mbox{while} \quad \frac{d\beta (X)}{ dX}
\sim \beta^2\,,
$$
 and $H-Z\sim\beta^2$. The expansion of the first two
supercharges in Eq.  (\ref{L8}) starts from $O(\beta )$, while that of the
last
two supercharges starts from $O(\beta^0 )$. To the leading order
\beq
Q_1& = &  i \psi_{1} \frac{\partial}{ \partial X}
- \psi_2\,  \beta (X)  \,,
  \quad
Q_2 =  i \psi_{2} \frac{\partial}{ \partial X}+
 \psi_1\, \beta (X)  \,,
   \nonumber \\ [0.2cm]
Q_3& = &   \psi_{4} \,, \qquad
Q_4 =   - \psi_{3} \,,
\eeq
which identically coincides with the nonrelativistic limit of (\ref{rsvv}),
provided $\beta (X)$ is identified with $W' (x)$ and $Z$ with 1/2,
see also  (\ref{dopseven}). The relativistic corrections
are different, however.
In particular, in the mirror-symmetric realization
$Q_3$ and $Q_4$ have corrections $O(\beta )$ while in the
realization (\ref{rsvv}) the first correction is  $O(\beta^2 )$.

\section{The Geometrical Meaning of Reduced
K\"ahler-Related  Realizations -- a Local
Reduction}

\subsection{Generalities}

Above we presented the SQM$_{2,Z}$ algebra (\ref{L8})
without explaining how we arrived at it.
In fact, it has a geometric origin.
Some  mysterious features of the mirror-symmetric realization  
(\ref{L8})
could be easily understood  since (locally) it is
obtained as a reduction of the standard representation of SQM$_{2}$,
which is obtained, in turn, by dimensional reduction
of the Wess-Zumino model. Moreover, the very defining property of
the ``mirror" symmetry finds a geometrical interpretation.

Let us remind that to obtain the standard quantum mechanics with four
supercharges one takes the Wess-Zumino model in four dimensions
(where the minimal number of supercharges is four)
and reduces it to one (time) dimension.
In this way one arrives at a supersymmetric quantal system
with the target space parametrized by one {\em complex} variable.
The existence of four supercharges implies that
it must be the K\"ahler space. Thus, in the standard case the number
of the (real) boson variables is two, while the number of
(real) fermion variables is four, so that Eq. (\ref{msone})
is satisfied.

Suppose we have the  algebra (\ref{one}),
and the operator $P$  commutes with the supercharges.
Then one can restrict oneself to the states that are the
eigenfunctions of the operator $P$, for example,
one can restrict to the eigenfunctions with the zero eigenvalue of $P$.
If the operator $P$ is a derivative along the  bosonic coordinate,
the reduced theory will have more fermionic degrees of freedom than
twice  the number of the  bosonic degrees of freedom.
To the best of our knowledge, this is the only way to construct
realizations of the centrally extended $SQM$
with $\nu_B <\nu_F/2$ --  by a reduction of the bosonic coordinates
in the
algebra (\ref{one}), through restriction to
the states which are annihilated by $P$.

Physically, this procedure can be interpreted as follows
\cite{3}. Consider a supersymmetric ${\cal N}=2$
system built of two
(interacting) components. This system can be described by two bosonic
coordinates, corresponding to the motion of the center
of mass of the system, plus the dynamics of an internal coordinate.
The motion of the center of mass can be factored out
in the general form. What remains is a reduced system with
$\nu_B=1$ and $\nu_F = 4$.

\subsection{The geometric formulation}

The standard solutions of the  SQM$_2$ equations
are constructed from the following geometrical data:
a complex manifold $M$ equipped with the  K\"ahler metric $G$ and a  
closed
holomorphic 1-differential on $M$ (the latter is needed for SQM$_{2Z}$),
\beq
\Omega=\Omega_n dz^n\,,\quad \partial \Omega =0\,,\quad \bar\partial 
\Omega
=0\,.
\label{Aone}
\eeq

Now we will outline the standard solutions (without central charges)
mainly with the purpose of introducing our
notation. We then  show how the central charges appear.

As well-known, the standard ${\cal N}=2$ superalgebra
is generated by four supercharges
\beq
Q_{\psi }&=&
i\partial +\bar\Omega = i\psi^m\frac{\partial}{\partial z^m}
+\bar\psi^{\bar m}  \bar\Omega_{\bar m}\, ,
\nonumber\\[0.2cm]
\bar Q_{\bar\psi}&=& i\bar\partial +\Omega = i\bar\psi^{\bar
m}\frac{\partial}{\partial \bar z^{\bar m}} -\psi^{ m}
\Omega_{ m}\, ,
\nonumber\\[0.2cm]
Q_{\chi }&=&
\left( i \partial +\bar\Omega\right)^\dagger =
i G^{\bar m n}\chi_n
\left(\frac{\partial}{\partial \bar z^{\bar m}} -\Gamma_{\bar m\bar 
n}^{\bar p}
\bar\psi^{\bar n} \bar\chi_{\bar p}
\right) + G^{\bar mn}\bar\chi_{\bar m} \Omega_n\, ,
\nonumber\\[0.2cm]
\bar Q_{\bar \chi }&=&
\left(  i\bar\partial + \Omega\right)^\dagger =
i G^{ m \bar n}\bar\chi_{\bar n}
\left(\frac{\partial}{\partial   z^{  m}} -\Gamma_{  m  n}^{  p}
\psi^{n} \chi_{  p}
\right)     -  G^{\bar mn}\chi_{n} \bar\Omega_{\bar m}\, ,
\label{Athree}
\eeq
where
$$
\psi^m = dz^m\,,\quad  \bar\psi^{\bar m} = d\bar z^{\bar m}\,,\quad
\chi_m =\iota \frac{\partial}{\partial   z^{  m}}\,,
\quad
\bar\chi_{\bar m} =\iota \frac{\partial}{\partial   \bar z^{\bar  m}}\,.
$$
Moreover, $\iota$ stands for the operation of the contraction 
with the vector field,
 $\Gamma_{  m  n}^{  p}$ is the Christoffel symbol corresponding to
the metric $G$, 
$$ 
\Gamma_{  m  n}^{  p} = G^{\bar\ell p}\partial_m G_{\bar\ell n}
= G^{\bar\ell p}\partial_n G_{\bar\ell m}\, ,\qquad
\bar\Gamma_{\bar m\bar 
n}^{\bar p} = G^{\bar p \ell }\partial_{\bar m} G_{\bar n\ell }
= G^{\bar p \ell }\partial_{\bar n} G_{\bar m \ell },
$$
and $\psi$, $\chi$
satisfy the following anticommutation relations:
\beq
\{\psi^m , \chi_n\} =\delta^m_n\,,\quad \{\bar\psi^{\bar m} , 
\bar\chi_{\bar
n}\}
=\delta^{\bar m}_{\bar n}\,.
\eeq
All other anticommutators are trivial.

Our basic idea, from which everything else  derives, is
as follows.  We
suppose that the manifold $M$
 has the U(1) Lie group of diffeomorphysms that preserves
metric, complex structure, and
$\Omega$.
 The Lie algebra of this group is represented by the vector field
 $V$,
\beq
V= V^m\frac{\partial}{\partial   z^{  m}} + \bar V^{\bar
m}\frac{\partial}{\partial
 \bar z^{\bar  m}}\,.
\eeq
The above conditions imply that
\beq
\Omega_m V^m ={\rm const}\,.
\eeq

In fact, below we will use a specific {\em ansatz}
for the metric,\footnote{In the case at hand a generic metric is reducible to
this {\em ansatz} by a coordinate transformation.} 
in which the metric $G_{\bar m n}$
will depend only on the combinations $z+\bar z$. Moreover,
$V^m$ will be a purely imaginary constant.

With the vector field switched on, one can obtain
the centrally extended algebra (provided
the standard supercharges (\ref{Athree}) are modified, see below),
\beq
\{ Q_{\psi }^V, Q_{\chi }^V\} &=& H+P_V\,,\quad
\{ \bar Q_{\bar \psi }^V, \bar Q_{\bar \chi }^V\} =H-P_V\,,
\nonumber\\[0.2cm]
\{ Q_{\psi }^V, \bar Q_{\bar \chi }^V\} &=&\bar  Z\,,\qquad
\{ \bar Q_{\bar \psi }^V,  Q_{\chi }^V \} = Z\,,
\eeq
where 
$H$ is the Hamiltonian,
\beq
H&=&-G^{\bar mn}\bar\partial_{\bar m}\partial_n  +G^{\bar nn}
\bar\Omega_{\bar n}\Omega_n + G_{\bar nn}
\bar V^{\bar n}V^n \nonumber\\[0.2cm]
&-& \left(\partial_m G^{\bar nn}\right)
\psi^m\chi_n \,\, \bar\partial_{\bar n}
-\left(\bar\partial_{\bar m} G^{\bar nn}\right)
\bar\psi^{\bar m}\bar\chi_{\bar n}\,\,
\partial_{ n}\nonumber\\[0.2cm]
&+& i\bar\Gamma^{\bar n}_{\bar p\bar m}\bar V^{\bar p}\bar\psi^{\bar m}
\bar\chi_{\bar n} - i\Gamma^p_{nq} V^n\psi^q\chi_p
\nonumber\\[0.2cm]
&-& \left(\partial_m\bar\partial_{\bar n}
G^{\bar p n}\right) \psi^m\chi_n\, \bar\psi^{\bar n}\bar\chi_{\bar
p}\nonumber\\[0.2cm]
&+& i\left[\partial_m
\left( G^{\bar m n}\Omega_n \right)\right] \psi^m\bar\chi_{\bar m}
-i \left[\bar\partial_{\bar m}
\left( G^{\bar n m}\bar\Omega_{\bar n}
\right)\right]\bar\psi^{\bar m}\chi_{n}\,,
\label{Hami}
\eeq
$P_V$ is the Lie derivative along $V$,
\beq
P_V = -i V^n \partial_n  -i  
\bar V^{\bar n} \bar\partial_{\bar n}  
-i\left(\partial_n\, V^m \psi^n\chi_m + \bar\partial_{\bar n}\, 
\bar V^{\bar m}
\bar\psi^{\bar n}\bar\chi_{\bar m}
\right)
\,,
\label{Ptwo}
\eeq
 and
$$
\partial_m\equiv \frac{\partial }{\partial z ^m}\,,\qquad 
\bar\partial_{\bar m}\equiv\frac{\partial }{\partial \bar z ^{\bar m}}\,.
$$
Finally, 
\beq
Z=2\,\,  \Omega_m V^m\,.
\eeq
Note that the last term in Eq. (\ref{Ptwo}) is irrelevant since 
our {\em ansatz} implies that  $\partial_n\, V^m =0$.

To obtain the Hamiltonian (\ref{Hami}) of the centrally extended
problem  one must modify the above supercharges by
adding vector-field-dependent terms, namely,
\beq
Q_{\psi }^V&=&
 i \psi^m\frac{\partial}{\partial z^m}
+\bar\psi^{\bar m}  \bar\Omega_{\bar m} -G_{m\bar n}\psi^m\bar 
V^{\bar n}\,
,
\nonumber\\[0.2cm]
\bar Q_{\bar\psi}^V&=& i \bar\psi^{\bar
m}\frac{\partial}{\partial \bar z^{\bar m}} -\psi^{ m}
\Omega_{ m} + G_{\bar m n}\bar\psi^{ \bar m} V^{ n} \, ,
\nonumber\\[0.2cm]
Q_{\chi }^V&=& i
G^{\bar m n}\chi_n
\left(\frac{\partial}{\partial \bar z^{\bar m}} -\Gamma_{\bar m\bar 
n}^{\bar
p}
\bar\psi^{\bar n} \bar\chi_{\bar p}
\right) + G^{\bar{m}n} \bar{\chi}_{\bar{m}}\Omega_n -
\chi_m V^m\, ,
\nonumber\\[0.2cm]
\bar Q_{\bar \chi }^V&=& i
G^{ m \bar n}\bar\chi_{\bar n}
\left(\frac{\partial}{\partial   z^{  m}} -\Gamma_{  m  n}^{  p}
\psi^{n} \chi_{  p}
\right) -
G^{\bar mn}\chi_n\bar\Omega_{\bar m} +
\bar\chi_{\bar m} \bar V^{\bar m}\, .
\label{Bthree}
\eeq

We are very close to our final goal. In order to obtain ${\cal N}=2$
centrally extended
superalgebra (\ref{two}) we have to make three steps.
First, we restrict ourselves to the sector of states on which $P_V =0$.
This is always possible since $P_V$ commutes with the supercharges
(provided that $\Omega$ is a constant).
This is just {the procedure of dynamical reduction}
we want to suggest. It reduces the number of the bosonic degrees of 
freedom
leaving that of the fermionic degrees of freedom intact. The second step 
is a similarity (quasigauge) transformation.
Finally, the third step is obvious:
it is the passage from the complex representation of the supercharges 
and $\psi$'s and $\chi$'s to real supercharges and $\psi$'s.
In the next subsection we will demonstrate how the procedure works
in a particular example.

\subsection{SQM$_{2Z}$   as a U(1) Reduction
of the Standard (K\"ahler) SQM$_2$}

Let us consider 
a concrete example. Assume $M$ to be a manifold of the cylinder
topology,  with the noncompact coordinate $X$, the
compact coordinate $Y$ and the complex structure
$z=X+iY$.  Take the K\"ahler metric in the following form:
\beq
ds^2= g^2(X)\left( dX^2 + dY^2 \right)\,,
\label{46}
\eeq
and 
\beq
\Omega =-i\, \omega dz\,,\qquad V=v\frac{\partial}{\partial Y}
=i\, v\left(\frac{\partial}{\partial z}-\frac{\partial}{\partial \bar z}
\right)\,,
\eeq
where $\omega$ and $v$ are real constants.

Substituting this particular metric and the corresponding Christoffel
symbols in
the general formulae
(\ref{Bthree}),  and setting $P_V=0$ (i.e. $\partial /\partial Y = 0$) we
get
\beq
Q_{\psi }^V&=&
 \frac{i}{2}\,\psi\frac{\partial}{\partial X}
+i\, \bar\psi\,  \omega +i\, g^2(X)\psi \, v\, ,
\nonumber\\[0.2cm]
\bar Q_{\bar\psi}^V&=& \frac{i}{2}\,\bar\psi
\frac{\partial}{\partial X} +i\, \psi
\omega +i\, g^2(X)\bar\psi \, v \, ,
\nonumber\\[0.2cm]
Q_{\chi }^V&=&  \frac{i}{2}\,
\frac{1}{g^2(X)}\chi
\left(\frac{\partial}{\partial  X} -2 \, \frac{\partial \ln g(X)}{\partial
X}\,
\bar\psi \bar\chi
\right) - \frac{i}{g^2(X)}\, \bar\chi\,\omega  -i\, \chi \, v\, ,
\nonumber\\[0.2cm]
\bar Q_{\bar \chi }^V&=&\frac{i}{2}\,
\frac{1}{g^2(X)}\bar\chi
\left(\frac{\partial}{\partial   X } -2 \, \frac{\partial \ln g(X)}{\partial
X}\,
\psi \chi
\right) +\frac{i}{g^2(X)}\, \chi\omega + i\, \bar\chi v\, .
\label{Bfour}
\eeq

Our task is to arrive at a symmetric expression for the first and the 
second
pair of the
supercharges. To this end we make a
  trick of  a similarity (gauge) transformation. We conjugate all operators
by
$
\exp\{-(\psi\chi +\bar\psi\bar \chi)\ln g \}\,,
$
namely,
\beq
\psi &\to& e^{-(\psi\chi +\bar\psi\bar \chi)\ln g(X) }\, \psi\,  e^{(\psi\chi
+\bar\psi\bar \chi)\ln g(X)}= \frac{1}{g(X)} \,\psi\,,
\nonumber\\[0.2cm]
\bar\psi &\to& e^{-(\psi\chi +\bar\psi\bar \chi)\ln g(X) }\,\bar\psi\, 
e^{(\psi\chi +\bar\psi\bar \chi)\ln g(X)}= \frac{1}{g(X)} \,\bar\psi\,,
\nonumber\\[0.2cm]
\chi &\to& e^{-(\psi\chi +\bar\psi\bar \chi)\ln g (X)}\, \chi \,  e^{(\psi\chi
+\bar\psi\bar \chi)\ln g(X)}= {g(X)} \,\chi \,,
\nonumber\\[0.2cm]
\bar \chi &\to& e^{-(\psi\chi +\bar\psi\bar \chi)\ln g(X) }\, \bar\chi \, 
e^{(\psi\chi +\bar\psi\bar \chi)\ln g(X)}= {g(X)} \,\bar\chi \,,
\eeq
and
\beq
\frac{\partial}{\partial X} \to e^{-(\psi\chi +\bar\psi\bar \chi)\ln g (X)}\,
\frac{\partial}{\partial X}\,   e^{(\psi\chi +\bar\psi\bar \chi)\ln g(X)} =
\frac{\partial}{\partial X} +(\psi\chi + \bar\psi\bar \chi)\,
 \frac{\partial \ln g(X)}{\partial X}\,.
\eeq

After the similarity transformations, the supercharges (\ref{Bfour})
become
\beq
Q_{\psi }^V&=&
 \frac{i}{2 g(X)} \, \psi \left[\frac{\partial}{\partial X}
+\bar\psi\bar\chi \frac{\partial g(X)}{\partial X}\right]
+\frac{i}{g(X)}\, \bar\psi\,  \omega +i\, g(X)\psi \, v\, ,
\nonumber\\[0.2cm]
\bar Q_{\bar\psi}^V&=& \frac{i}{2g(X)} \, \bar\psi
\left[ \frac{\partial}{\partial X} +\psi\chi\,  \frac{\partial g(X)}{\partial
X}\right]
 +\frac{i}{g(X)}\, \psi\omega +i\, g(X)\bar\psi \, v \, ,
\nonumber\\[0.2cm]
Q_{\chi }^V&=&  \frac{i}{2 g(X)}\chi
\left[\frac{\partial}{\partial  X} +\bar\chi\bar\psi  \, \frac{\partial \ln
g(X)}{\partial X}
\right] - \frac{i}{g(X)}\, \bar\chi\,\omega  -i\, g(X) \chi \, v\, ,
\nonumber\\[0.2cm]
\bar Q_{\bar \chi }^V&=&\frac{i}{2 g(X)}\bar\chi
\left[ \frac{\partial}{\partial   X } +\chi\psi \, \frac{\partial \ln g(X)}{\partial
X}
\right] -\frac{i}{g(X)}\, \chi\omega - i\,g(X) \bar\chi v\, .
\label{Bfive}
\eeq

It is not difficult to check
that  the four supercharges (\ref{Bfive}), and those of (\ref{L8}),
are in one-to-one correspondence.
To this end one forms the following combinations:
\beq
Q_1 = \frac{1}{2}\left( Q^V_\psi + Q^V_\chi - \bar Q^V_{\bar\psi} -
\bar Q^V_{\bar\chi}\right)\, ,
\nonumber\\[0.2cm]
Q_2 = \frac{1}{2i}\left( Q^V_\psi - Q^V_\chi  - \bar Q^V_{\bar\psi} +
\bar Q^V_{\bar\chi}
\right)\, ,
\nonumber\\[0.2cm]
Q_3 = \frac{1}{2}\left( Q^V_\psi + Q^V_\chi + \bar Q^V_{\bar\psi} +
\bar Q^V_{\bar\chi} 
\right)\, ,
\nonumber\\[0.2cm]
Q_4 = \frac{1}{2i}\left( Q^V_\psi - Q^V_\chi + \bar Q^V_{\bar\psi} -
\bar Q^V_{\bar\chi} 
\right)\, ,
\label{Pone}
\eeq
and
\beq
\psi_1 = \frac{1}{\sqrt{2}}\left(\psi -\bar\psi +\chi -\bar\chi \right)\, ,
\nonumber\\[0.2cm]
\psi_2 = \frac{1}{i\sqrt{2}}\left(\psi -\bar\psi -\chi +\bar\chi \right)\, ,
\nonumber\\[0.2cm]
\psi_3 = \frac{1}{\sqrt{2}}\left(\psi +\bar\psi +\chi + \bar\chi \right)\, ,
\nonumber\\[0.2cm]
\psi_4 = \frac{1}{i\sqrt{2}}\left(\psi +\bar\psi -\chi -\bar\chi \right)\, .
\label{Pthree}
\eeq

Then one  expresses $Q_{1,2,3,4}$ in terms of $\psi_{1,2,3,4}$,
performs an additional similarity transformation
$$
\frac{\partial}{\partial X} \to e^{(1/2)\ln g (X)}\,
\frac{\partial}{\partial X}\,   e^{-(1/2) \ln g(X)} =
\frac{\partial}{\partial X} -\frac{1}{2}
 \frac{\partial \ln g(X)}{\partial X}\,,
$$
and passes from the variable $X$ to a new variable $X'$,
\beq
X\to X' =2\sqrt{2}\int g(X) dX\,.
\eeq
In this way one   arrives at four supercharges equivalent to (\ref{L8}).
As was already mentioned,  at $v=\omega = 0$ we observe an unexpected
(from the  geometric
standpoint) overextension of the $R$ symmetry, see Sec. 2.2.

In light of our geometric understanding,
the mirror symmetry discussed in Sec. 3.2, 
presents the equivalence of quantum mechanics obtained
by the reduction from two perfectly different manifolds
(\ref{46}), with distinct scale factors, see Fig. 1. 

\begin{figure}[h]
\epsfxsize=11cm
\begin{center}
\leavevmode
\epsfbox{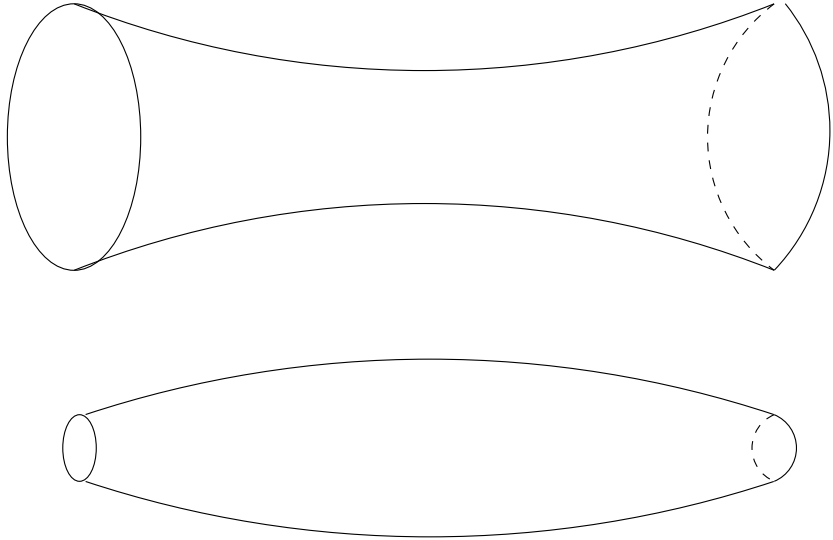}
\caption{Quantum mechanics obtained through reduction 
from these two manifolds are identical.}
\end{center}
\end{figure}

\subsection{Global vs. local}

So far it was assumed that the coordinate $X$ is noncompact.
Let us now compactify it, i.e. proceed to the toroidal topology.
Equation (\ref{L9}) implies then that the period $\Pi$ of the
function $f(X)$ defined in Eq. (\ref{Lsix}) vanishes
if we start from  any well-defined two-dimensional
K\"ahler manifold. However, for the existence of SQM$_{2,Z}$
it is sufficient to assume that (i) $f(X)$ is single-valued on $M$;
(ii) Equation (\ref{L9}) is valid locally, i.e. the integral
$\int f(X) dX $ need not be single-valued. Thus, if we take the $X$
space to be
 $S_1$ and the function $f$ with a nonvanishing period, we arrive at
SQM$_{2,Z}$ which cannot  be  globally
obtained by  reduction from any two-dimensional K\"ahler manifold.
(At the moment we do not know whether this solution can be obtained
 by reduction from a higher-dimensional K\"ahler manifold.)

\section{Conclusions}

We have constructed, via dynamical reduction, a rather strange looking
${\cal
N}=2$ centrally extended quantum mechanics with $\nu_B< \nu_F /2$.
Our geometrical construction partially explains
the properties of the system obtained. However, not everything is 
explained.
In particular, the origin of the mirror symmetry and the
overextended  $R$ symmetry are not
understood.

As follows from our work, the problem of the general classification
of extended superalgebras
(with the standard or even overextended $R$ symmetries)
in the form (\ref{cerepre}) is an open question. The solution of
this problem may well lead to new interesting geometrical structures 
encoded
in the coefficients $e,\omega,\zeta$, see Eq. (\ref{cerepre}),
which could be considered as a generalization of the
K\"ahler geometry.

\vspace{0.3cm}

The authors acknowledge numerous discussions with A. Smilga and 
 A. Vainshtein. 
We are very grateful to S. Bellucci, 
Jan van Holten, Evgeny Ivanov, S. Krivonos, and A. Pashnev
for useful communications.
 
This work  was 
supported by DOE grant DE-FG02-94ER408.
The work of A.L. was supported in part
by RFFI grant 01-01-00548, Support for Scientific Schools grant
00-15-96-557, and INTAS grant 99-590.

\end{document}